\documentclass[twocolumn]{autart}    

\usepackage{mathrsfs}
\usepackage{color}
\usepackage{amsfonts}
\usepackage{subfigure}
\usepackage{booktabs}
\usepackage{balance}
\usepackage{graphicx} 
\usepackage{float} 
\usepackage{enumerate}
\usepackage{bm}
\usepackage{makecell}
\usepackage{epstopdf}
\usepackage{mathtools}
\usepackage{bbm}
\usepackage{dsfont}
\usepackage{pifont}
\usepackage{cite}
\usepackage{balance}
\usepackage[T1]{fontenc}
\usepackage{soul}
\usepackage[normalem]{ulem}

\newtheorem{remark}{Remark}

\newtheorem{theorem}{Theorem}
\newtheorem{lemma}{Lemma}

\newtheorem{problem}{Problem}
\newtheorem{definition}{Definition}
\newtheorem{proposition}{Proposition}

\newcommand\R{\mathbb{R}}
\newcommand\pr{\operatorname{Pr}}
\newcommand\var{\operatorname{Var}}
\newcommand\N{\mathbb{N}}
\newcommand\E{\mathbb{E}}

\newcommand\hs{\mathrm{H}_{\mathrm{s}}}
\newcommand\hd{\mathrm{H}_{\mathrm{d}}}
\newcommand\kl{D_{\mathcal{KL}}}
\newcommand\dd{\mathrm{d}}

\usepackage[monochrome]{xcolor}

\begin{document}
\begin{frontmatter}

    \title{Probabilistic Predictability of  Stochastic Dynamical Systems\thanksref{footnoteinfo}} 

    \thanks[footnoteinfo]{Preliminary results have been published in the 61-th IEEE Conference on Decision and Control \cite{xuPredictabilityStochasticDynamical2022a}.}

    \author[Shanghai]{Tao Xu}\ead{Zerken@sjtu.edu.cn},
    \author[Shanghai]{Yushan Li}\ead{yushan\_li@sjtu.edu.cn},
    \author[Shanghai]{Jianping He}\ead{jphe@sjtu.edu.cn}

    \address[Shanghai]{Shanghai Jiao Tong University, China}


    \begin{abstract}                          
        To assess the quality of a probabilistic prediction for stochastic dynamical systems (SDSs), scoring rules assign a numerical score based on the predictive distribution and the measured state.
        In this paper, we propose an $\epsilon$-logarithm score that generalizes the celebrated logarithm score by considering a neighborhood with radius $\epsilon$.
        We characterize the probabilistic predictability of an SDS by optimizing the expected score over the space of probability measures. We show how the probabilistic predictability is quantitatively determined by the neighborhood radius, the differential entropies of process noises, and the system dimension. \textcolor{blue}{Given any predictor, we provide approximations for the expected score with an error of scale $\mathcal{O}(\epsilon)$.}
        In addition to the expected score, we also analyze the asymptotic behaviors of the score on individual trajectories. Specifically, we prove that the score on a trajectory can converge to the expected score when the process noises are independent and identically distributed. Moreover, the convergence speed against the trajectory length $T$ is of scale $\mathcal{O}(T^{-\frac{1}{2}})$ in the sense of probability.
        Finally, numerical examples are given to elaborate the results.
    \end{abstract}

\end{frontmatter}

\section{Introduction}
\subsection{Background}
Noises are inevitable in dynamical systems, resulting in prediction uncertainties for future state trajectories. A probabilistic predictor predicts the target by a distribution rather than a single point, which can inherently quantify the prediction uncertainties. Therefore, probabilistic prediction for stochastic dynamical systems (SDSs) has attracted a surge of recent attention \cite{landgrafProbabilisticPredictionMethods2023}.

To measure the quality of a probabilistic prediction, scoring rules assign a numerical score based on the predictive distribution and the realized outcome \cite{gneitingStrictlyProperScoring2007a,gneitingProbabilisticForecasting2014}. A scoring rule is called proper if the expected score is maximized when the predictive distribution equals the ground truth, which motivates the predictor to be unbiased in predicting the true distribution. One of the most celebrated proper scoring rules is the logarithm score, which assigns the score as the logarithm value of the predictive \textcolor{blue}{probability density function (pdf) or probability mass function (pmf)} at the realized outcome.


\subsection{Motivations}
\textcolor{blue}{Due to the potential nonlinear SDS dynamics and non-Gaussian system noises, the target pdfs are typically multimodal in a variety of scenarios. For state estimation tasks with multimodal observation likelihoods, particle filters can provide multimodal estimations\cite{vaswaniParticleFilteringLargedimensional2008}; for edge detection tasks in computer vision, the regression-by-classification approach can model multimodal distributions under aleatoric uncertainties \cite{xiongHingeWassersteinEstimatingMultimodal2024}.} By taking into account the neighborhood with tunable radius $\epsilon$, we propose an $\epsilon$-logarithm score in this paper. It degenerates into the traditional logarithm score when $\epsilon$ equals $0$.

A popular line of research is to design algorithms to probabilistically predict the state trajectories of SDSs, aiming for feasibility guarantee \cite{kouvaritakisExplicitUseProbabilistic2010}, better robustness \cite{sauderProbabilisticRobustDesign2019}, higher accuracy \cite{roelofseAccurateEfficientApproach2023}, etc. Given a scoring rule, the probabilistic predictability of an SDS can be naturally characterized by the optimal expected score. It may greatly boost the efficiency of designing predictors if we have a deeper understanding of the predictability of an SDS, e.g., what system features directly affect predictability and which one possesses the largest weight.

Although the expected score is theoretically appealing in characterizing the system's predictability, practically evaluating its value requires a sufficient amount of repeated samples for averaging. However, the samples generated from a typical SDS prediction scenario are usually temporal (a trajectory of states) rather than spatial (repeated samplings for the state at a fixed time step). While the average of spatial score samplings converges to the expectation as ensured by the law of large numbers, there is no simple guarantee for the average of temporal score samplings. Given any single trajectory generated from an SDS, under what condition can the temporal averaged score converge? Will it converge to the expected score? How fast the convergence can be?


\subsection{Contributions}
The main contributions are summarized as follows.
\begin{itemize}
    \item (Metric) We propose an $\epsilon$-logarithm score that generalizes the celebrated logarithm score by considering a neighborhood with radius $\epsilon$. When $\epsilon$ equals $0$, the proposed score will degenerate to the logarithm score. \textcolor{blue}{Given any predictor, we approximate the expected $\epsilon$-logarithm score with the error of scale $\mathcal{O}(\epsilon)$.}
    \item (Optimality) We characterize the probabilistic predictability of an SDS by the optimal expected $\epsilon$-logarithm score, regardless of specific prediction algorithms. It quantitatively strengthens our understanding of how a system's predictability is jointly determined by the neighborhood radius, the differential entropies of process noises and the state dimension.
    \item (Convergence) We analyze the asymptotic convergence behaviors of the proposed score on any single trajectory generated from an SDS. It is proved that the score can converge to the expected score when the process noises are independent and identically distributed. Furthermore, the convergence speed against the trajectory length $T$ is guaranteed to be of scale $\mathcal{O}(T^{-\frac{1}{2}})$ in the sense of probability.
\end{itemize}

The remainder of this paper is organized as follows. Section II reviews the related works. Sec. III  formulates the problems of interest. Sec. IV characterize the system's predictability by evaluating the optimal expected score. Sec. V introduces a partition-based method to approximate expected scores and analyzes the asymptotic convergence behavior. Simulations are shown in Sec. VI.

\section{Related Works}

A lot of insightful works contribute to the predictability analysis of deterministic dynamical systems. Lorenz considered prediction performance as the growing rate of initial state uncertainty, then defined predictability as the asymptotic exponential growing rate of initial prediction error \cite{lorenzPredictabilityProblemPartly1996}. Motivated by this idea, some famous indexes such as Lyapunov exponent and Kolmogorov-Sinai entropy were proposed to characterize the predictability of dynamical systems, see a review of these indexes in \cite{boffettaPredictabilityWayCharacterize2002}. These early predictability analyses have found wide applications in the climatology fields \cite{kalnayAtmosphericModelingData2003}. However, these works do not take noises or state measurements into consideration, and thus can not be directly applied to characterize the predictability of SDSs.

Research on the predictability analysis of discrete-state SDSs mainly bifurcates into two directions. A body of research treats the predictability of SDS from an information-theoretic perspective without first evaluating the prediction performance, thus a lot of entropy-based predictability metrics were proposed. The entropy of stochastic process is defined as the joint entropy in \cite{biondiMaximizingEntropyMarkov2014}, based on which optimal prediction performance analysis and unpredictable system designs were presented in \cite{savasEntropyMaximizationPartially2022a}. Another line of research steers the complicated evaluation of prediction performance by approximation techniques. In \cite{songLimitsPredictabilityHuman2010}, an upper bound of the accurate prediction probability is derived based on standard Fano's inequality. This bound is applied to the study of large-scale urban vehicular mobility \cite{liLimitsPredictabilityLargeScale2014}.

Research on the predictability analysis of continuous-state SDS is relatively less than the discrete ones. In the field of climate forecasting, the predictability of an SDS is defined as the distance between a predicted distribution and climatological distribution based on entropy, relative entropy and mutual information \cite{delsolePredictabilityInformationTheory2004,delsolePredictabilityInformationTheory2005,delsolePredictabilityRecentInsights2007}. In the field of state estimation, some concern the predictability as the effect of model mismatch on the steady solution of the Kalman filter \cite{byrnesPredictabilityUnpredictabilityKalman1991}, some study the predictability by evaluating the worst-case mean square error prediction performance of the Kalman filter \cite{yasiniWorstCasePredictionPerformance2018}. Recently, an unpredictable design of SDS was developed in \cite{liUnpredictableTrajectoryDesign2020}, which formulated an optimization problem with $\epsilon$-accurate prediction probability as the objective.

However, existing works on predictability analysis of SDSs mainly serve for point predictions, and it remains open and challenging to analyze the predictability under a probabilistic prediction framework.

\section{Preliminaries and Problem Formulation}
\subsection{Preliminaries}
\paragraph*{Notations} We denote random variables in bold fonts to distinguish them from constant variables. \textcolor{blue}{Given a random variable $\mathbf{x}$, we denote its pdf (or pmf) as $p_{\mathbf{x}}(\cdot)$. We also denote a sequence $\{(\cdot)_k\}_{k=1}^T$ by $(\cdot)_{1:T}$, and $(\cdot)_{1:0}$ is defined as an empty set.} \textcolor{blue}{For a vector $v$, we use $v_{(i)}$ to deonte the $i$th element of $v$. Given a set $A$, the indicator function $\mathbb{I}_A(x)$ equals $1$ if $x\in A$, otherwise equals $0$. We use $*$ and $*\!/\!*$ to denote the convolution and deconvolution operators, respectively. $\log(0)$'s value is set to $0$.}

\paragraph*{Entropy} The Shannon entropy of a discrete random variable $\mathbf{x}$ with alphabet $\mathcal{X}$ and pmf $p_\mathbf{x}$ is $\hs(p_{\mathbf{x}}):=-\sum_{x \in \mathcal{X}} p_\mathbf{x}(x) \log p_\mathbf{x}(x)$.
The differential entropy of a continuous random variable $\mathbf{x}$ with support $\mathcal{X}$ and pdf $p_\mathbf{x}$ is $\hd({p_\mathbf{x}}):=-\int_{x \in \mathcal{X}} p_\mathbf{x}(x) \log p_\mathbf{x}(x) \dd x$.
\textcolor{blue}{The KL-divergence measures how a distribution $p_{\mathbf{x}}$ is different from another distribution $\hat{p}_{\mathbf{x}}$, $\kl(p_{\mathbf{x}} || \hat{p}_{\mathbf{x}}):=\mathbb{E}_{x \in p_{\mathbf{x}}}\log(\frac{p_{\mathbf{x}}(x)}{\hat{p}_{\mathbf{x}}(x)})$.} \textcolor{blue}{The cross entropy of a distribution $p_{\mathbf{x}}$ relateive to another distribution $\hat{p}_{\mathbf{x}}$ is $\mathrm{H}(p_{\mathbf{x}} || \hat{p}_{\mathbf{x}})\!:= - \mathbb{E}_{x \in p_{\mathbf{x}}} \log(\hat{p}_{\mathbf{x}}(x))$.}

\paragraph*{Probabilistic prediction} The problem of probabilistic prediction can be generally formulated as follows. Suppose a random variable $\mathbf{x}$ takes value on $\mathcal{X}$ with distribution $p_\mathbf{x}$, a probabilistic predictor predicts it by a distribution $\hat{p}_{\mathbf{x}} \in \mathcal{P}$, where $\mathcal{P}$ is a family of distributions over $\mathcal{X}$. When the value of $\mathbf{x}$ is materialized as $x$, a scoring rule,
\begin{equation}
    S(\hat{p}_{\mathbf{x}}, x): \mathcal{P}\times \mathcal{X} \to \R,
\end{equation}
assigns a numerical score $S(\hat{p}_{\mathbf{x}}, x)$ to measure the quality of the predictive distribution $\hat{p}_{\mathbf{x}}$ on the realized value $x$. The expected scoring rule of $S$,
usually sharing the same operator $S:\mathcal{P}\times \mathcal{P}  \to \R$ but different operands, is
\begin{equation}
    S(\hat{p}_{\mathbf{x}}, p_{\mathbf{x}}) := \E_{x\sim p_{\mathbf{x}}} S(\hat{p}_{\mathbf{x}}, x).
\end{equation}
A scoring rule $S$ is proper with respect to the prediction space $\mathcal{P}$ if \textcolor{blue}{$S(\hat{p}_{\mathbf{x}}, p_{\mathbf{x}}) \leq S(p_{\mathbf{x}}, p_{\mathbf{x}})$} holds for all $\hat{p}_{\mathbf{x}}, p_{\mathbf{x}} \in \mathcal{P}$. It is strictly proper if the equality holds only when $\hat{p}_{\mathbf{x}}= p_{\mathbf{x}}$.
For example, the logarithm score, $\mathcal{L}(\hat{p}_{\mathbf{x}}, x):= \log \hat{p}_{\mathbf{x}}(x)$, is most celebrated for being essentially the only local proper scoring rule up to equivalence \cite{parryProperLocalScoring2012a}; the linear score,
$\operatorname{LinS}(\hat{p}_{\mathbf{x}}, x):= \hat{p}_{\mathbf{x}}(x)$, is not a proper scoring rule, despite its intuitive appeal in both theory and practice \cite{gneitingStrictlyProperScoring2007a}.

\subsection{System and Predictor Model}
Consider a discrete-time stochastic dynamical system, denoted by $\Phi$,
\begin{equation}\label{eq:sds}
    \Phi: \mathbf{x}_{k+1} =f(\mathbf{x}_k) + \mathbf{w}_k,
\end{equation}
where $\mathbf{x}_k\in \mathbb{R}^{d_x}$ is the system state and \textcolor{blue}{the dynamics $f: \R^{d_x} \to \R^{d_x}$ is continuous. Let $\mathcal{P}$ be the family of Lebesgue integrable pdfs over $\mathbb{R}^{d_x}$, the process noises $\{\mathbf{w}_k\}_{k=1}^\infty$ with pdf $p_{\mathbf{w}_k} \in \mathcal{P}$} are not necessarily required to be independent and identically distributed (i.i.d).

A probabilistic predictor keeps observing the states of $\Phi$ and predicting the conditional pdfs of future states. \textcolor{blue}{Specifically, given previous observations $x_{1:k}$ at time step $k$, a probabilistic prediction is denoted as
\begin{equation}\label{eq:predictor}
    x_{1:k} \to \hat{p}_{\mathbf{x}_{k+1} \mid \mathbf{x}_{1:k}}(\cdot \mid x_{1:k}) \in \mathcal{P}.
\end{equation}
After the value of $\mathbf{x}_{k+1}$ is realized as $x_{k+1}$ from the true pdf $p_{\mathbf{x}_{k+1} \mid \mathbf{x}_{1:k}}(\cdot \mid x_{1:k}) \in \mathcal{P}$, the one-step prediction performance is measured by a scoring rule $S$ as $S(\hat{p}_{\mathbf{x}_{k+1} \mid \mathbf{x}_{1:k}}(\cdot \mid x_{1:k}), x_{k+1})$.}

\subsection{Problems of Interest}
The logarithm score assesses the prediction performance by the predictive pdf at the exact observed value without considering its neighborhood. Taking a neighborhood with radius $\epsilon$ into account, we generalize the logarithm score into an $\epsilon$-logarithm score as follows.
\begin{definition}[$\epsilon$-logarithm score]
    Given a neighborhood radius $\epsilon \geq 0$, a random variable $\mathbf{x}$ to be predicted, the $\epsilon$-logarithm score evaluates the quality of a predictive distribution $\hat{p}_{\mathbf{x}}\in\mathcal{P}$ on a realized outcome $x$ by
    \begin{equation}
        \mathcal{L}_\epsilon(\hat{p}_{\mathbf{x}}, x) := \left\{ \begin{array}{lr}
            \log \hat{p}_{\mathbf{x}}(x)                                           & \epsilon = 0, \\
            \log \int_{\|s-x\|_\infty \leq \epsilon} \hat{p}_{\mathbf{x}}(s) \dd s & \epsilon > 0,
        \end{array}\right.
    \end{equation}
    and the expected $\epsilon$-logarithm score is
    \begin{equation}
        \mathcal{L}_\epsilon(\hat{p}_{\mathbf{x}}, p_\mathbf{x}) := \E_{x\sim p_{\mathbf{x}}}\, \mathcal{L}_\epsilon(\hat{p}_{\mathbf{x}}, x).
    \end{equation}
\end{definition}

While the $\epsilon$-logarithm score $\mathcal{L}_\epsilon(p_{\mathbf{x}}, x)$ scores a one-step prediction, we can naturally extend this definition to the trajectory prediction of SDSs.
\begin{definition}[$\epsilon$-logarithm score for SDSs]
    Given $\epsilon\geq 0$, a state trajectory $x_{1:T}$ generated from an SDS $\Phi$, the $\epsilon$-logarithm score for a probabilistic predictor $\hat{p}$ on this trajectory is the average of one-step scores, i.e.,
    \begin{equation}
        \bar{\mathcal{L}}_\epsilon(\hat{p}_{\mathbf{x}_{1:T}}, x_{1:T}) \!:=\! \frac{1}{T}\!\sum_{k=1}^{T} \mathcal{L}_\epsilon\left(\hat{p}_{\mathbf{x}_k \mid \mathbf{x}_{1:k-1}}(\cdot \mid x_{1:k-1}), x_k\right),
    \end{equation}
    where $\hat{p}_{\mathbf{x}_k \mid \mathbf{x}_{1:k-1}} \in \mathcal{P}$ for $k=1,\ldots,T$. The expected $\epsilon$-logarithm score is denoted as
    \begin{equation}
        \bar{\mathcal{L}}_\epsilon(\hat{p}_{\mathbf{x}_{1:T}}, p_{\mathbf{x}_{1:T}}) := \E_{x_{1:T}}\bar{\mathcal{L}}_\epsilon(\hat{p}_{\mathbf{x}_{1:T}}, x_{1:T}).
    \end{equation}

\end{definition}

\begin{problem}
Evaluate the expected $\epsilon$-logarithm score and characterize the probabilistic predictability of an SDS by optimizing the expected $\epsilon$-logarithm score, \textcolor{blue}{i.e.,
    \begin{equation}\label{eq:opt-expected}
        \max_{\hat{p}_{\mathbf{x}_k \mid \mathbf{x}_{1:k-1}} \in \mathcal{P} \;\forall k \in [1,T]} \bar{\mathcal{L}}_\epsilon(\hat{p}_{\mathbf{x}_{1:T}}, p_{\mathbf{x}_{1:T}}).
    \end{equation}}
\end{problem}

Though the expected $\epsilon$-logarithm score is theoretically appealing, practically evaluating it requires many samples for averaging. However, the data under a typical SDS prediction scenario is usually a state trajectory rather than repeated samplings for trajectories.
\begin{problem}
Given any state trajectory $x_{1:\infty}$ generated from $\Phi$, does the $\epsilon$-logarithm score $\bar{\mathcal{L}}_\epsilon(\hat{p}_{\mathbf{x}_{1:T}}, x_{1:T})$ converge as $T$ approaches infinity? If it does converge, what is the converged value and how fast the convergence is?
\end{problem}

\section{Probabilistic Predictability of SDSs}
\textcolor{blue}{Let $\mathbf{u}_{\epsilon}$ be a random variable subjected to a uniform distribution over the $\epsilon$-ball $\{x\in\R^{d_x} \mid \|x\|_\infty \leq \epsilon\}$, and we denote its pdf as $p_{\mathbf{u}_\epsilon}$. Convolving each pdf in $\mathcal{P}$ by $p_{\mathbf{u}_\epsilon}$, we aggregate the outcomes as the set $\mathscr{F}_\epsilon = \left\{\tilde{p} \mid \tilde{p} = \hat{p} * p_{\mathbf{u}_\epsilon}, \hat{p} \in \mathcal{P}\right\}$. Then, we obtain the probabilistic predictability of SDSs by solving \eqref{eq:opt-expected}.
    \begin{theorem}[Probabilistic predictability]\label{thm:predictability}
        Given a neighborhood radius $\epsilon \geq 0$ and a trajectory length $T \geq 1$, \\
        i) the optimal expected $\epsilon$-logarithm score for predicting the trajectory $\mathbf{x}_{1:T}$ of an SDS $\Phi$ is
        \begin{equation}\label{eq:predictability}
            \begin{aligned}
                  & \max_{\hat{p}_{\mathbf{x}_k \mid \mathbf{x}_{1:k-1}} \in \mathcal{P} \;\forall k \in [1,T]}  \bar{\mathcal{L}}_\epsilon(\hat{p}_{\mathbf{x}_{1:T}}, p_{\mathbf{x}_{1:T}}) \\
                = & d_x \log(2\epsilon) - \frac{1}{T} \mathrm{H}(p_{\mathbf{x}_{1:T}}||\tilde{p}^\star_{\mathbf{x}_{1:T}}),
            \end{aligned}
        \end{equation}
        where $\tilde{p}^\star_{\mathbf{x}_{1:T}}$ satisfies the following equations $\forall\,k \in [1,T]$,
        \begin{equation}\label{eq:opt-pp}
            \tilde{p}^\star_{\mathbf{x}_k \mid \mathbf{x}_{1:k-1}} = \arg\min_{\tilde{p}\in\mathscr{F}_\epsilon} \mathrm{H}(p_{\mathbf{x}_k \mid \mathbf{x}_{1:k-1}} || \tilde{p}).
        \end{equation}\\
        ii) The optimal predictor $p^\star_{\mathbf{x}_k \mid \mathbf{x}_{1:k-1}}$ attaining the predictability is the deconvolution of $\tilde{p}^\star_{\mathbf{x}_k \mid \mathbf{x}_{1:k-1}}$ by $p_{\mathbf{u}_\epsilon}$, i.e.,
        \begin{equation}\label{eq:deconvolution}
            p^\star_{\mathbf{x}_k \mid \mathbf{x}_{1:k-1}} = \tilde{p}^\star_{\mathbf{x}_k \mid \mathbf{x}_{1:k-1}} *\!\!/\!\!* p_{\mathbf{u}_\epsilon}.
        \end{equation}
    \end{theorem}
    \begin{pf}
        When $\epsilon = 0$, $p_{\mathbf{u}_\epsilon}$ is a Dirac function and $\mathcal{F}_\epsilon = \mathcal{P}$, then \eqref{eq:predictability} and \eqref{eq:opt-pp} are derived because the value of $\log(0)$ is set to be $0$. Next, the deconvolution \eqref{eq:deconvolution} admits that $p^\star = p$. When $\epsilon > 0$, the expected $\epsilon$-logarithm score at any step $k \in [1,T]$ is
        \begin{equation*}
            \begin{aligned}
                                      & \mathcal{L}_\epsilon\left(\hat{p}_{\mathbf{x}_k \mid \mathbf{x}_{1:k-1}}(\cdot \mid x_{1:k-1}), p_{\mathbf{x}_k \mid \mathbf{x}_{1:k-1}}(\cdot \mid x_{1:k-1})\right) \\
                \overset{(i)}{=}      & \int_{x \in \R^{d_x}} q(x) \log \left( \int_{\|s-x\|_\infty\leq\epsilon} \hat{q}(s) \dd s \right) \dd x                                                               \\
                =                     & d_x\log(2\epsilon) + \int_{x \in \R^{d_x}} q(x) \log \left( \frac{\int_{\|s-x\|_\infty\leq\epsilon} \hat{q}(s) \dd s}{(2\epsilon)^{d_x}} \right) \dd x                \\
                \overset{(ii)}{=}     & d_x\log(2\epsilon) + \int_{x \in \R^{d_x}} q(x) \log\left[\hat{q} * p_{\mathbf{u}_\epsilon}(x)\right]  \dd x                                                          \\
                \overset{(iii)}{\leq} & d_x\log(2\epsilon) - \mathrm{H}(q || \tilde{q}^\star),
            \end{aligned}
        \end{equation*}
        where $(i)$ follows by denoting $p_{\mathbf{x}_k \mid \mathbf{x}_{1:k-1}}(\cdot \mid x_{1:k-1})$ as $q(\cdot)$ and $\hat{p}_{\mathbf{x}_k \mid \mathbf{x}_{1:k-1}}(\cdot \mid x_{1:k-1})$ as $\hat{q}(\cdot)$; (ii) follows from the definition of convolution and (iii) holds by denoting $\tilde{p}^\star_{\mathbf{x}_k \mid \mathbf{x}_{1:k-1}}(\cdot \mid x_{1:k-1})$ as $\tilde{q}^\star$ and the definition in \eqref{eq:opt-pp}.
        Therefore, $\bar{\mathcal{L}}_\epsilon(\hat{p}_{\mathbf{x}_{1:T}}, p_{\mathbf{x}_{1:T}})$ can be maximized as follows
        \begin{equation*}
            \begin{aligned}
                                    & \mathbb{E}_{x_{1:T}}\frac{1}{T}\!\sum_{k=1}^{T} \mathcal{L}_\epsilon\left(\hat{p}_{\mathbf{x}_k \mid \mathbf{x}_{1:k-1}}(\cdot \mid x_{1:k-1}), x_k\right)                                                                                          \\
                =                   & \frac{1}{T}\!\!\sum_{k=1}^{T} \mathbb{E}_{x_{1:k\!-\!1}}\mathcal{L}_\epsilon\!\left(\hat{p}_{\mathbf{x}_k \mid \mathbf{x}_{1:k\!-\!1}}\!(\cdot \!\mid\! x_{1:k-1}), p_{\mathbf{x}_k \mid \mathbf{x}_{1:k\!-\!1}}\!(\cdot \!\mid\! x_{1:k-1})\right) \\
                \overset{(i)}{\leq} & \frac{1}{T}\!\!\sum_{k=1}^{T} \mathbb{E}_{x_{1:k\!-\!1}} \left[ d_x\log(2\epsilon) -\right.                                                                                                                                                         \\
                                    & \left.\mathrm{H}\left(p_{\mathbf{x}_k \mid \mathbf{x}_{1:k-1}}(\cdot \mid x_{1:k-1}) || \tilde{p}^\star_{\mathbf{x}_k \mid \mathbf{x}_{1:k-1}}(\cdot \mid x_{1:k-1}) \right) \right]                                                                \\
                \overset{(ii)}{=}   & d_x \log(2\epsilon) - \frac{1}{T}\mathrm{H}(p_{\mathbf{x}_{1:T}} || \tilde{p}^\star_{\mathbf{x}_{1:T}}),
            \end{aligned}
        \end{equation*}
        where $(i)$ follows from the previous one-step inequality, and the equality holds when \eqref{eq:opt-pp} is satisfied; $(ii)$ holds according to the chain rules for joint entropy and relative entropy \cite[pp. 22-24]{coverElementsInformationTheory2006}. Given $\tilde{p}^\star_{\mathbf{x}_k \mid \mathbf{x}_{1:k-1}} \in \mathcal{F}_\epsilon$, there is $p^\star_{\mathbf{x}_k \mid \mathbf{x}_{1:k-1}} \in \mathcal{P}$ such that \eqref{eq:deconvolution} holds, which is the optimal predictor that attains the probalistic predictability.
    \end{pf}
    \begin{remark}
        One can get an upper bound for \eqref{eq:predictability}:
        \begin{equation}\label{eq:predictability_upper_bound}
            \begin{aligned}
                 & d_x \!\log(2\epsilon) \!-\! \frac{1}{T} \mathrm{H}(p_{\mathbf{x}_{1:T}}||\tilde{p}^\star_{\mathbf{x}_{1:T}}) \\ \leq &d_x \!\log(2\epsilon) \!-\! \frac{1}{T} \hd(p_{\mathbf{x}_{1:T}}),
            \end{aligned}
        \end{equation}
        where the equality holds if and only if $p_{\mathbf{x}_k \mid \mathbf{x}_{1:k-1}} \in \mathscr{F}_\epsilon \forall k \in [1,T]$, then $\tilde{p}^\star_{\mathbf{x}_k \mid \mathbf{x}_{1:k-1}} = p_{\mathbf{x}_k \mid \mathbf{x}_{1:k-1}}$ and the optimal predictor $p^\star_{\mathbf{x}_k \mid \mathbf{x}_{1:k-1}}$ is the deconvolution $p_{\mathbf{x}_k \mid \mathbf{x}_{1:k-1}} *\!\!/\!\!* p_{\mathbf{u}_\epsilon}$. As for a general $\epsilon \in \R_{+}$, the density deconvolution may not exist. For example, $\forall \tilde{p} \in \mathscr{F}_\epsilon$, $\exists \hat{p}\in\mathcal{P}$ such that $\tilde{p} = \hat{p} * p_{\mathbf{u}_\epsilon}$. When $\hd(p_{\mathbf{u}_\epsilon}) > \hd(p_{\mathbf{x}_k \mid \mathbf{x}_{1:k-1}})$, there is $\hd(\tilde{p}) = \hd(\hat{p} * p_{\mathbf{u}_\epsilon}) \geq \hd(p_{\mathbf{u}_\epsilon})> \hd(p_{\mathbf{x}_k \mid \mathbf{x}_{1:k-1}})$, and we have $p_{\mathbf{x}_k \mid \mathbf{x}_{1:k-1}} \notin \mathscr{F}_\epsilon$. In other words, a necessary condition for the equality of \eqref{eq:predictability_upper_bound} to hold is $\hd(p_{\mathbf{u}_\epsilon}) \leq \hd(p_{\mathbf{x}_k \mid \mathbf{x}_{1:k-1}})$. Extremely, as $\epsilon\to\infty$ there is $\hd(p_{\mathbf{u}_\epsilon}) \to \infty$, this necessary condition will not hold.
    \end{remark}}

\section{Approximation and Convergence Analysis}
In this section, we first provide approximations to the expected $\epsilon$-logarithm score with an error of scale $\mathcal{O}(\epsilon)$. Next, while $\bar{\mathcal{L}}_\epsilon(\hat{p}_{\mathbf{x}_{1:T}}, p_{\mathbf{x}_{1:T}})$ considers the expected performance over all possible trajectories, we also analyze the asymptotic behaviors of the $\bar{\mathcal{L}}_\epsilon(\hat{p}_{\mathbf{x}_{1:T}}, x_{1:T})$ on any single trajectory. In particular, we characterize the convergence speed for the SDSs with i.i.d process noises.

Typically, there is no explicit expression for $\mathcal{L}_\epsilon(\hat{p}_{\mathbf{x}_{1:T}}, p_{\mathbf{x}_{1:T}})$ under general pdfs. To handle this problem, we first utilize a partition-based method to formally evaluate $\mathcal{L}_\epsilon$.
\textcolor{blue}{\begin{definition}[Unifrom grid partition]
        Given $l \in \R_{+}$, a uniform grid partition $\Sigma^l$ is a set that divides $\R^{d_x}$ into disjoint cubes with edge lengths equaling $l$. Specifically, we denote
        \begin{equation*}
            \Sigma^{\ell} := \left\{A_{v}^{\ell}\right\}_{v\in \mathbb{Z}^{d_x}}, \quad \Theta_{\Sigma^\ell}(x) := \sum_{v\in\mathbb{Z}^{d_x}} v\cdot\mathbb{I}_{A_v^\ell}(x),
        \end{equation*}
        where $A_{v}^{\ell} = \{a \in \mathbb{R}^{d_x} \mid a_{(i)} \in \left(v_{(i)}\ell, (v_{(i)}+1)\ell\right]\}$ is a cube in $\mathbb{R}^{d_x}$ with size $\ell$ and index $v$. $\forall x \in \R^{d_x}$, $\Theta_{\Sigma^\ell}(x)$ is the cube index of $x$ under partition $\Sigma^\ell$.
    \end{definition}}
\textcolor{blue}{Given a uniform grid partition $\Sigma^\ell$, a pdf $\hat{p}_{\mathbf{x}}$ can be approximated by a pmf $\hat{p}_{\mathbf{x}}^{\Sigma^\ell}$, where
\begin{equation}
    \hat{p}_{\mathbf{x}}^{\Sigma^\ell}(v) := \int\nolimits_{A_v^\ell} \hat{p}_{\mathbf{x}}(s)\dd s, \forall v \in \mathbb{Z}^{d_x}.
\end{equation}
Then we define the $\Sigma^\ell$-logarithm score as $\mathcal{L}_{\Sigma^\ell}(\hat{p}_{\mathbf{x}}, x) := \log \hat{p}_{\mathbf{x}}^{\Sigma^\ell}(\Theta_{\Sigma^\ell}(x))$, and the expected $\Sigma^\ell$-logarithm score is defined as $\mathcal{L}_{\Sigma^\ell}(\hat{p}_{\mathbf{x}}, p_{\mathbf{x}}) := \E_x\mathcal{L}_{\Sigma^\ell}(\hat{p}_{\mathbf{x}}, x)$.}

$\mathcal{L}_{\Sigma^\ell}$ can be evaluated as the negative cross entropy.
\begin{proposition}[Evaluation of $\mathcal{L}_{\Sigma^\ell}$]\label{thm:evl_finite}
    Given a random variable $\mathbf{x}$, a partition $\Sigma^\ell$ on $\mathbb{R}^{d_x}$ and a probabilistic predictor $\hat{p}_\mathbf{x}$, the expected $\Sigma^\ell$-logarithm score is
    \begin{equation}\label{eq:evl_finite}
        \mathcal{L}_{\Sigma^\ell}(\hat{p}_{\mathbf{x}}, {p}_{\mathbf{x}}) = -\mathrm{H}(p_{\mathbf{x}}^{\Sigma^\ell}||\hat{p}_{\mathbf{x}}^{\Sigma^\ell}).
    \end{equation}
\end{proposition}


To evaluate $\mathcal{L}_\epsilon$, we first develop the following lemma to address an inequality relationship between $\mathcal{L}_\epsilon$ and $\mathcal{L}_{\Sigma^\ell}$.
\begin{lemma}\label{lem:ineq}
    Given a neighborhood radius $\epsilon \geq 0$, a pdf $p_{\mathbf{x}}\in\mathcal{P}$ and a predictor $\hat{p}_\mathbf{x}\in\mathcal{P}$, $\mathcal{L}_\epsilon(\hat{p}_\mathbf{x}, p_\mathbf{x})$ is bounded by
    \textcolor{blue}{\begin{equation}\label{eq:bounded_ineq}
            \mathcal{L}_{\Sigma^\epsilon}(\hat{p}_\mathbf{x}, p_\mathbf{x})           \leq \mathcal{L}_\epsilon(\hat{p}_\mathbf{x}, p_\mathbf{x})
            \leq  \max\limits_{\ell > 0} \mathcal{L}_{\Sigma^\ell}(\hat{p}_\mathbf{x}, p_\mathbf{x}).
        \end{equation}}
\end{lemma}
\begin{pf}
    Please see Appendix \ref{app:lem:ineq}.
\end{pf}
This lemma provides a coarse way to bound the $\epsilon$-logarithm score by $\Sigma^\ell$-logarithm score. It helps to guarantee the existence of a special partition $\Sigma^\star$ that transforms the evaluation of $\epsilon$-logarithm score to evaluating $\Sigma^\star$-logarithm score.

\begin{theorem}[Evaluation of $\mathcal{L}_\epsilon$] \label{thm:existence}
    \textcolor{blue}{Given $\epsilon\geq 0$, a pdf $p_{\mathbf{x}} \in \mathcal{P}$ and a predictor $\hat{p}_\mathbf{x} \in \mathcal{P}$, there exists a uniform grid partition} $\Sigma^\star$ such that
    \begin{equation}\label{eq:formal}
        \mathcal{L}_\epsilon(\hat{p}_{\mathbf{x}}, p_\mathbf{x}) = -\mathrm{H}(p_{\mathbf{x}}^{\Sigma^\star}||\hat{p}_{\mathbf{x}}^{\Sigma^\star}).
    \end{equation}
\end{theorem}
\begin{pf}
    Please see Appendix \ref{app:thm:existence}.
\end{pf}
Exploiting the partition-based formal evaluation, we can approximate $\mathcal{L}_\epsilon$ when pdfs are continuous. \textcolor{blue}{Let $\mathcal{P}_c$ denote the space of continuous pdfs over $\R^{d_x}$, we have the following lemma.}
\begin{lemma}[Approximation of $\mathcal{L}_\epsilon(\hat{p}_\mathbf{x}, p_\mathbf{x})$]\label{lem:approx_onestepLog}
    Given $\epsilon \geq 0$, \textcolor{blue}{a pdf $p_\mathbf{x} \in \mathcal{P}_c$ and a probabilistic predictor $\hat{p}_\mathbf{x}\in \mathcal{P}_c$}, the expected $\epsilon$-logarithm score $\mathcal{L}_\epsilon(\hat{p}_\mathbf{x}, p_\mathbf{x})$ is approximated as
    \begin{equation*}
        \left|\mathcal{L}_\epsilon(\hat{p}_\mathbf{x}, p_\mathbf{x}) - d_x\log(2\epsilon)+\mathrm{H}(p_{\mathbf{x}}||\hat{p}_{\mathbf{x}})\right| = \mathcal{O}(\epsilon).
    \end{equation*}
\end{lemma}
\begin{pf}
    Please see Appendix \ref{app:lem:approx_onestepLog}.
\end{pf}

Next, we can extend this one-step approximation result to the SDS trajectory.
\begin{theorem}[Approximation of $\bar{\mathcal{L}}_\epsilon(\hat{p}_{\mathbf{x}_{1:T}}, p_{\mathbf{x}_{1:T}})$]\label{thm:approx_elog_traj}
    Given $\epsilon \geq 0$, a trajectory $\mathbf{x}_{1:T}$ with \textcolor{blue}{continuous pdf $p_{\mathbf{x}_{1:T}}$ and a predictor $\hat{p}_{\mathbf{x}_k \mid \mathbf{x}_{1:k-1}} \in \mathcal{P}_c$} at each step $k$, the expected $\epsilon$-logarithm score $\bar{\mathcal{L}}_\epsilon(\hat{p}_{\mathbf{x}_{1:T}}, p_{\mathbf{x}_{1:T}})$ is approximated as
    \begin{equation*}
        \begin{aligned}
            \left|\bar{\mathcal{L}}_\epsilon(\hat{p}_{\mathbf{x}_{1:T}}, p_{\mathbf{x}_{1:T}}) \!-\! d_x\log(2\epsilon)\!+\!\frac{1}{T}\mathrm{H}(p_{\mathbf{x}_{1:T}} || \hat{p}_{\mathbf{x}_{1:T}})\right| \!=\! O(\epsilon).
        \end{aligned}
    \end{equation*}
\end{theorem}
\begin{pf}
    Please see Appendix \ref{app:thm:approx_elog_traj}.
\end{pf}


Particularly when $\hat{p}_{\mathbf{x}_{1:T}} = p_{\mathbf{x}_{1:T}}$, the convergence of $\bar{\mathcal{L}}_\epsilon$ is mainly determined by the convergence of $\frac{1}{T} \hd(\mathbf{x}_{1:T})$, which is the entropy rate of the stochastic process $\{\mathbf{x}_k\}_{k=1}^\infty$. However, the entropy rate is not guaranteed \textcolor{blue}{to always exist} when the process noises are not i.i.d \cite[p. 74]{coverElementsInformationTheory2006}.
Therefore, to analyze the convergence of the $\epsilon$-logarithm score for a given state trajectory, one should focus on the SDSs with i.i.d process noises. \textcolor{blue}{Additionally, when the system dynamics are known to the predictor, the predictive pdf for state $\mathbf{x}_k$ can be reduced to a predictive pdf $\hat{p}_{\mathbf{w}} \in \mathcal{P}$ for the noise $\mathbf{w}$, i.e.,
    \begin{equation}\label{eq:predictor_iid}
        \hat{p}_{\mathbf{x}_k \mid \mathbf{x}_{1:k-1}}(x_k \mid x_{1:k-1}) = \hat{p}_{\mathbf{w}}(x_k - f(x_{k-1})).
    \end{equation}}
\begin{theorem}\label{thm:converge}
    Given $\epsilon \geq 0$, a state trajectory $x_{1:T}$ of an SDS subjected to i.i.d process noises with pdf $p_\mathbf{w}\in\mathcal{P}$ and a predictor $\hat{p}_{\mathbf{x}_{1:T}}$ satisfying \eqref{eq:predictor_iid},\\
    \textcolor{blue}{i) the expected $\epsilon$-logarithm score is $\mathcal{L}_\epsilon(\hat{p}_\mathbf{w}, p_\mathbf{w})$, i.e.,
        \begin{equation*}
            \bar{\mathcal{L}}_\epsilon(\hat{p}_{\mathbf{x}_{1:T}}, p_{\mathbf{x}_{1:T}}) = \mathcal{L}_\epsilon(\hat{p}_{\mathbf{w}}, p_\mathbf{w}).
        \end{equation*}\\
        ii) As $T$ approaches infinity, the $\epsilon$-logarithm score on any single trajectory converges to the expectation, i.e.,
        \begin{equation*}
            \lim\limits_{T\to\infty} \bar{\mathcal{L}}_\epsilon(\hat{p}_{\mathbf{x}_{1:T}}, x_{1:T}) \overset{P}{=} \bar{\mathcal{L}}_\epsilon(\hat{p}_{\mathbf{x}_{1:T}}, p_{\mathbf{x}_{1:T}}).
        \end{equation*}}\\
    iii) Moreover, if $\E_w \mathcal{L}_\epsilon(\hat{p}_\mathbf{w}, w)^2 < \infty$, the convergence speed is $\mathcal{O}_p(\frac{1}{\sqrt{T}})$, i.e., $\forall \delta > 0$, there is
    \begin{equation*}
        \pr\left\{|\bar{\mathcal{L}}_\epsilon(\hat{p}_{\mathbf{x}_{1:T}}, x_{1:T}) - \bar{\mathcal{L}}_\epsilon(\hat{p}_{\mathbf{x}_{1:T}}, p_{\mathbf{x}_{1:T}})| \geq \delta \right\} = \mathcal{O}(\frac{1}{\sqrt{T}}).
    \end{equation*}
\end{theorem}
\begin{pf}
    Please see Appendix \ref{app:thm:converge}.
\end{pf}
\begin{remark}
    \textcolor{blue}{In practice, there is no need to use the sample average of a large number of trajectories to approximate $\bar{\mathcal{L}}_\epsilon(\hat{p}_{\mathbf{x}_{1:T}}, p_{\mathbf{x}_{1:T}})$ when the SDS has i.i.d process noises. Specifically, as guaranteed by Theorem \ref{thm:converge}, calculating the $\epsilon$-logarithm score on any single trajectory $x_{1:T}$ will quickly converge to the expectation with the speed $\mathcal{O}_p(\frac{1}{\sqrt{T}})$.}
\end{remark}

\begin{figure*}[t]
    \centering
    \subfigure[$\epsilon$=0.25]{
        \includegraphics[width = 0.31\textwidth]{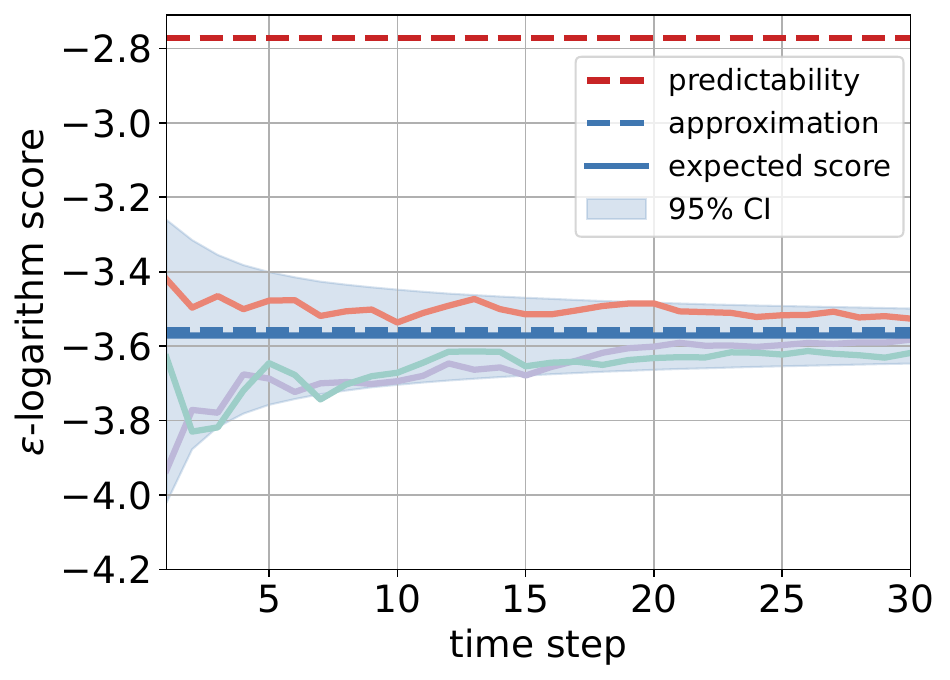}
        \label{sim:0.25}
    }
    \subfigure[$\epsilon$=0.50]{
        \includegraphics[width = 0.31\textwidth]{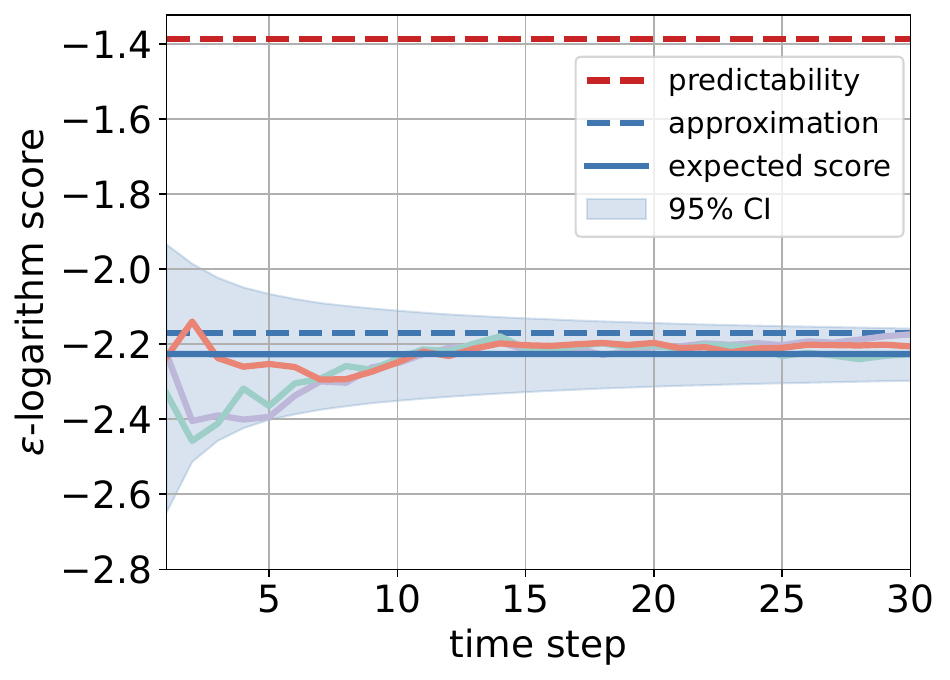}
        \label{sim:0.50}
    }
    \subfigure[$\epsilon$=0.75]{
        \includegraphics[width = 0.31\textwidth]{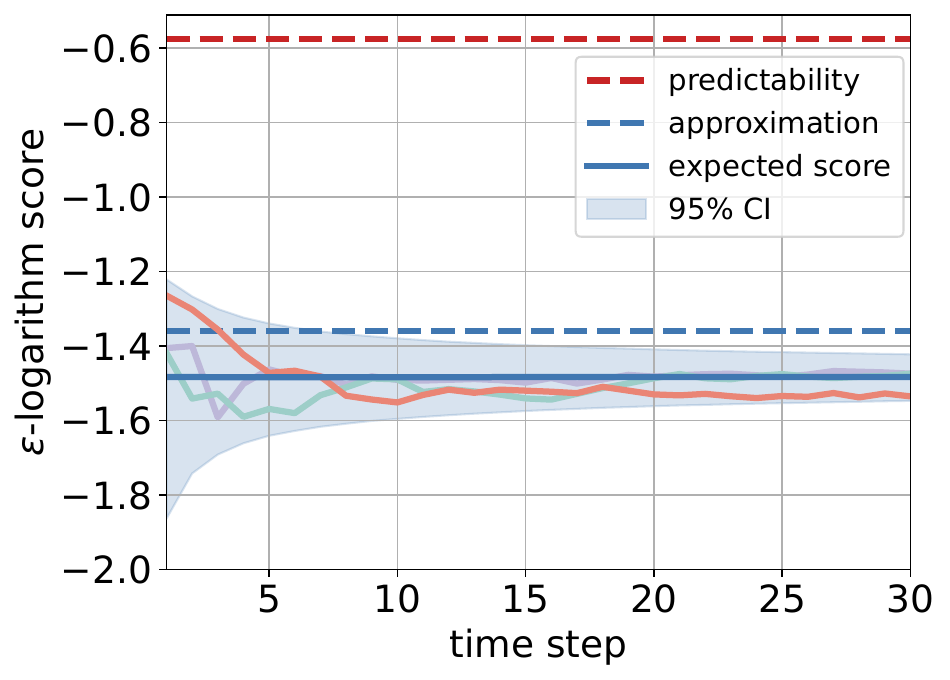}
        \label{sim:0.75}
    }
    \caption{$\epsilon$-logarithm scores \textcolor{blue}{$\bar{\mathcal{L}}_\epsilon(p_{\mathbf{x}_{1:T}}, x_{1:T}^{(n)})$ v.s. the time step $T$: given $\epsilon$ and trajectories $\{x_{1:T}^{(n)}\}_{n=1}^{10^5}$ of $\Phi$, i) score curves on three individual trajectories are randomly chosen for presentation; ii) the expected scores (blue solid line) and 95\% confidence intervals (blue transparent area) at each step are calculated from the scores on $100,000$ trajectories; iii) the predictability (red dotted line) is evaluated by Theorem \ref{thm:predictability} and the expected score's approximation (blue dotted line) is evaluated by Theorem \ref{thm:approx_elog_traj}.}}
    \label{sim:eps}
\end{figure*}

\section{Simulation}
\subsection{Simulation Setup}
\textcolor{blue}{We study a linear SDS subjected to i.i.d noises that are uniformly distributed over the cube $[-1,1]^2$,
\begin{equation*}
    \begin{array}{cc}
        \Phi: \begin{array}{lcl}
                  \mathbf{x}_{k+1} & = & \begin{pmatrix}
                                       1 & 1 \\
                                       0 & 1
                                   \end{pmatrix} \mathbf{x}_k+ \mathbf{w}_k,\quad \mathbf{w}_k \overset{\text{i.i.d}}{\sim} U([-1,1]^2).
              \end{array}
    \end{array}
\end{equation*}
Then, we randomly generate trajectories $\{x_{1:T}^{(n)}\}_{n=1}^{10^5}$ of $\Phi$ with length $T=30$ starting from a random initial state. The predictor $\hat{p}_{\mathbf{x}_{1:T}}$ satisfies the condition \eqref{eq:predictor_iid} in Theorem \ref{thm:converge}, and we let $\hat{p}_{\mathbf{w}}$ be a standard two-dimensional Gaussian distribution. We choose the neighborhood radius $\epsilon$ as $0.25, 0.5, 0.75$ respectively.
Next, we numerically compute the score $\bar{\mathcal{L}}_\epsilon(\hat{p}_{\mathbf{x}_{1:T}}, x_{1:T}^{(n)})$ for each trajectory. In Fig. \ref{sim:eps}, we use the scores' mean to evaluate the expected score and calculate $0.025$ and $0.975$ quantiles to form a 95\% confidence interval. According to Theorem \ref{thm:predictability}, the predictability of $\Phi$ can be explicitly calculated as $2\log(2\epsilon)-\log 4$, plotted as red dotted line. According to Theorem \ref{thm:approx_elog_traj}, the expected score can be approximated by $2\log(2\epsilon) - \log(2\pi) - \frac{1}{3}$, plotted as blue dotted line. Finally, we randomly choose three trajectories for each $\epsilon$ and plotted their score curves.}

\subsection{Results and Analysis}
\textcolor{blue}{As Fig. $\ref{sim:eps}$ shows, the distances between the blue solid lines (the real expected score) and the blue dotted lines (the approximated expected score) are indeed of scale $\mathcal{O}(\epsilon)$. Even for the extreme setting in Fig. \ref{sim:0.75}, where $\epsilon = 0.75$ is close to the radius $1$ of the noises' support $[-1,1]^2$, the approximation error is still around $0.1$. Nevertheless, a practically reasonable choice for the error tolerance $\epsilon$ should always be much smaller than the support of noises, otherwise the performances of different predictors would be indistinguishable. Therefore, our approximation in Theorem \ref{thm:approx_elog_traj} is effective.}

Next, Fig. $\ref{sim:eps}$ shows that all individual trajectories approach fast to the blue dotted lines in less than $20$ time steps. This quick convergence is ensured by Theorem \ref{thm:converge} in the sense of probability.
Benefiting from the quick convergence property of the score on individual trajectories, evaluating the expected score is easy to implement on one trajectory without the need for repeated samplings of different trajectories.

\section{Conclusion}
In this paper, we have proposed an $\epsilon$-logarithm score to assess the performance of probabilistic predictions in stochastic dynamical systems. We have evaluated the probabilistic predictability of an SDS by optimizing the expected score over the space of probability measures. It has allowed us to quantitatively analyze how the predictability of the system depends on the neighborhood radius, differential entropies of process noises, and system dimension. Additionally, we have provided approximations to the expected score for general non-optimal predictors. We have also analyzed the asymptotic convergence behavior of our score on any individual trajectory. It is proved that the score converges to the expected score when the process noises are independent and identically distributed, with a convergence speed of scale $\mathcal{O}_p(T^{-\frac{1}{2}})$ for the trajectory length $T$.

\bibliographystyle{ieeetr}        
\bibliography{ref_v2}           

\appendix

\section{Proof of Lemma \ref{lem:ineq}}\label{app:lem:ineq}
\textcolor{blue}{For the convenience of notation, we omit $\infty$ in the notation of $\|\cdot\|_\infty$ from now on.}

\textcolor{blue}{Guaranteed by the law of large numbers, given samples $\{x^r\}_{r=1}^\infty$ where $x^r \overset{i.i.d}{\sim} p_{\mathbf{x}}$, there is
\begin{equation*}
    \begin{aligned}
        \mathcal{L}_\epsilon(\hat{p}_{\mathbf{x}}, p_{\mathbf{x}})= & \int_{\mathbb{R}^{d_x}} p_\mathbf{x}(x)  \log \left(\int_{\|\hat{x}-x\|\leq\epsilon} \hat{p}_\mathbf{x}(\hat{x})  \dd \hat{x}\right) \dd x        \\
        =                                                           & \lim\limits_{R\to\infty} \frac{1}{R} \sum_{r=1}^{R} \log \left(\int_{\|\hat{x}-x^r\|\leq\epsilon}\hat{p}_\mathbf{x}(\hat{x})  \dd \hat{x}\right).
    \end{aligned}
\end{equation*}
Viewing the integration above as the expectation for an indicator random variable $\mathbf{1}_{\|\hat{\mathbf{x}}-x^r\|\leq\epsilon}$ (the indicator takes value $1$ when $\|\hat{\mathbf{x}}-x^r\|\leq\epsilon$ and takes value $0$ otherwise), it can be expressed as the asymptotic mean based on samples $\{\hat{x}^r_k\}_{k=1}^\infty$ where $\hat{x}^r_k \overset{i.i.d}{\sim} \hat{p}_{\mathbf{x}}$, i.e.,
\begin{equation*}
    \begin{aligned}
        \int_{\|\hat{x}-x^r\|\leq\epsilon}\!\hat{p}_\mathbf{x}(\hat{x})  \dd \hat{x} = & \mathbb{E}_{\mathbf{\hat{x}}} \mathbf{1}_{\|\hat{\mathbf{x}}-x^r\|\leq\epsilon} \!
        = \! \lim\limits_{K\to\infty}\frac{1}{K}\!\!\sum_{k=1}^{K} \mathcal{K}^{(k,r)}_\epsilon,
    \end{aligned}
\end{equation*}
where $\mathcal{K}^{(k,r)}_\epsilon = \mathbf{1}_{\|\hat{x}^r_k-x^r\|\leq\epsilon}$. It follows that
\begin{equation*}
    \mathcal{L}_\epsilon(\hat{p}_{\mathbf{x}}, p_{\mathbf{x}}) = \lim\limits_{R\to\infty} \frac{1}{R} \sum_{r=1}^{R} \log \left(\lim\limits_{K\to\infty}\frac{1}{K}\sum_{k=1}^{K} \mathcal{K}^{(k,r)}_\epsilon \right).
\end{equation*}
Similarly, the expected $\Sigma^\ell$-logarithm score is
\begin{equation*}
    \mathcal{L}_{\Sigma^\ell}(\hat{p}_{\mathbf{x}}, p_{\mathbf{x}}) = \lim\limits_{R\to\infty} \frac{1}{R} \sum_{r=1}^{R} \log \left(\lim\limits_{K\to\infty}\frac{1}{K}\sum_{k=1}^{K} \mathcal{K}^{(k,r)}_{\Sigma^\ell} \right),
\end{equation*}
where $\mathcal{K}^{(k,r)}_{\Sigma^\ell} = \mathbf{1}_{\Theta_{\Sigma^\ell}(\hat{x}^r_k) = \Theta_{\Sigma^\ell}(x^r)}$.}

\textcolor{blue}{For the left inequality, there is
\begin{equation*}
    \begin{aligned}
        \mathcal{K}^{(k,r)}_{\Sigma^\epsilon} = 1 \overset{(i)}{\Leftrightarrow} \Theta_{\Sigma^\epsilon}(\hat{x}_k^r) = \Theta_{\Sigma^\epsilon}(x^r)
         & \overset{(ii)}{\Rightarrow}  \|\hat{x}_k^r - x^r\| \leq \epsilon \\ &\overset{(iii)}{\Leftrightarrow} \mathcal{K}^{(k,r)}_\epsilon = 1,
    \end{aligned}
\end{equation*}
where $(i)$ and $(iii)$ follows from the definitions of $\mathcal{K}^{(k,r)}_{\Sigma^\epsilon}$ and $\mathcal{K}^{(k,r)}_\epsilon$ respectively; $(ii)$ holds because $\Theta_{\Sigma^\epsilon}(\hat{x}_k^r) = \Theta_{\Sigma^\epsilon}(x^r)$ indicates the existence of an $\epsilon$-sized cube $A\in\Sigma^\epsilon$ such that $\hat{x}_k^r, x^r \in A$, thus $\|\hat{x}_k^r - x^r\| \leq \epsilon$.
Therefore, given any $R, K\in\mathbb{R}_+$ and any possible samples $\left\{x^r, \{\hat{x}^r_k\}_{k=1}^K\right\}_{r=1}^R$, there is $\mathcal{K}^{(k,r)}_{\Sigma^\epsilon} \leq \mathcal{K}^{(k,r)}_\epsilon$. Thus
\begin{equation*}
    \frac{1}{R} \sum_{r=1}^{R} \log \left(\frac{1}{K}\sum_{k=1}^{K} \mathcal{K}^{(k,r)}_{\Sigma^\epsilon} \right) \leq \frac{1}{R} \sum_{r=1}^{R} \log \left(\frac{1}{K}\sum_{k=1}^{K} \mathcal{K}^{(k,r)}_\epsilon \right).
\end{equation*}
Letting $R, K \to \infty$, we have $\mathcal{L}_{\Sigma^\epsilon}(\hat{p}_\mathbf{x}, p_\mathbf{x}) \leq \mathcal{L}_\epsilon(\hat{p}_\mathbf{x}, p_\mathbf{x})$. Considering the extreme case where $\Sigma = \{\R^{d_x}\}$, given any $R, K\in\mathbb{Z}_+$ and samples $\left\{x^r, \{\hat{x}^r_k\}_{k=1}^K\right\}_{r=1}^R$, there is $\mathcal{K}^{(k,r)}_\Sigma \geq \mathcal{K}^{(k,r)}_\epsilon$. Thus
$$
    \mathcal{L}_\epsilon(\hat{p}_\mathbf{x}, p_\mathbf{x})\leq\max\limits_{\ell > 0} \mathcal{L}_{\Sigma^\ell}(\hat{p}_\mathbf{x}, p_\mathbf{x}).
$$}

\section{Proof of Theorem \ref{thm:existence}}\label{app:thm:existence}
\textcolor{blue}{Given pdfs $\hat{p}_\mathbf{x}, p_\mathbf{x}$ over $\R^{d_x}$, define function $\mathcal{F}: \R_{+} \to \R $ as $\mathcal{F}(\ell):=\mathcal{L}_{\Sigma^\ell}(\hat{p}_\mathbf{x}, p_\mathbf{x})$.
We first prove that $\mathcal{F}$ is continuous, i.e., for any $l > 0$, we need to prove that given any $\xi > 0$, there exists $\delta > 0$ s.t. $\forall \ell \in (l-\delta, l + \delta)$, $|\mathcal{F}(\ell) - \mathcal{F}(l)| < \xi$.  Since
\begin{equation*}
    \begin{aligned}
        \mathcal{F}(\ell) = & \lim_{N\to\infty} \sum_{\|v\| \leq N} p_{\mathbf{x}}^{\Sigma^\ell}(v) \log \left(\hat{p}_{\mathbf{x}}^{\Sigma^\ell}(v)\right),
    \end{aligned}
\end{equation*}
then $\forall \ell > 0$, $\exists N_{\xi}(\ell) \in \N$ such that
\begin{equation*}
    \left|\mathcal{F}(\ell) - \sum_{\|v\| \leq N_{\xi}(\ell)} p_{\mathbf{x}}^{\Sigma^\ell}(v) \log \left(\hat{p}_{\mathbf{x}}^{\Sigma^\ell}(v)\right)\right| < \frac{\xi}{4}.
\end{equation*}
Let $\bar{N} = \sup_{\kappa\in[0.9,1.1]}N_{\xi}(\kappa l)$, $\forall v$ with $\|v\|\leq \bar{N}$, one gets
\begin{equation*}
    \begin{aligned}
        \left|p_{\mathbf{x}}^{\Sigma^\ell}(v) - p_{\mathbf{x}}^{\Sigma^l}(v)\right|
        \leq & \left|\int_{A_v^\ell \Delta A_v^l} p_{\mathbf{x}}(x)\dd x\right|,
    \end{aligned}
\end{equation*}
where $\Delta$ denotes the symmetric difference between two measurable sets. Let $\mu(\cdot)$ denote the Lebesgue measure, there is $\lim_{\ell\to l}\mu(A_v^\ell \Delta A_v^l) = 0$. Thus \[\lim_{\ell\to l} \left|p_{\mathbf{x}}^{\Sigma^\ell}(v) - p_{\mathbf{x}}^{\Sigma^l}(v)\right| = 0,\] which means $p_{\mathbf{x}}^{\Sigma^\ell}(v)$ is continuous at $l$. Similarly, one has $\hat{p}_{\mathbf{x}}^{\Sigma^\ell}(v)$ is continuous at $l$. Then, there exists $\delta > 0$ such that when $|\ell - l| \leq \delta$ one has
\begin{equation*}
    \left|\sum_{\|v\| \leq \bar{N}}\!\! \left\{ p_{\mathbf{x}}^{\Sigma^\ell}(v) \log \left(\hat{p}_{\mathbf{x}}^{\Sigma^\ell}(v)\right) \!-\! p_{\mathbf{x}}^{\Sigma^l}(v) \log \left(\hat{p}_{\mathbf{x}}^{\Sigma^l}(v)\right) \right\}\right|  \!\leq\!  \frac{\xi}{2}.
\end{equation*}
Hence, for any $\ell$ such that $|\ell - l|\leq \min\{\delta, 0.1l\}$, there is
\begin{equation*}
    \begin{aligned}
         & \left|\mathcal{F}(\ell) \!\!-\!\!\! \sum_{\|v\| \leq \bar{N}}\!\! p_{\mathbf{x}}^{\Sigma^\ell}\!(v)\! \log\! \left(\hat{p}_{\mathbf{x}}^{\Sigma^\ell}(v)\right) \!\!+\!\!\!\! \sum_{\|v\| \leq \bar{N}}\!\! p_{\mathbf{x}}^{\Sigma^\ell}\!(v)\! \log\! \left(\hat{p}_{\mathbf{x}}^{\Sigma^\ell}(v)\right)\right. \\
         & \left.\!- \!\!\!\!\sum_{\|v\| \leq \bar{N}}\!\! p_{\mathbf{x}}^{\Sigma^l}\!(v)\! \log\! \left(\hat{p}_{\mathbf{x}}^{\Sigma^l}(v)\right)\!+\!\!\!\! \sum_{\|v\| \leq \bar{N}}\!\! p_{\mathbf{x}}^{\Sigma^l}\!(v)\! \log\! \left(\hat{p}_{\mathbf{x}}^{\Sigma^l}(v)\right)\!-\!\! \mathcal{F}(l) \right|           \\
         & \leq \frac{\xi}{4} + \frac{\xi}{2} + \frac{\xi}{4} = \xi,
    \end{aligned}
\end{equation*}
and the continuity of $\mathcal{F}$ is proved.
Next, as indicated by Lemma \ref{lem:ineq}, $\lim_{\ell \to \infty} \mathcal{F}(\ell) = 0 > \mathcal{L}_\epsilon(\hat{p}_{\mathbf{x}}, p_{\mathbf{x}})$. Since $\mathcal{F}$ is continuous, there exists $l_1 > \epsilon$ such that $\mathcal{F}(l_1) > \mathcal{L}_\epsilon(\hat{p}_{\mathbf{x}}, p_{\mathbf{x}})$. Applying the intermediate value theorem \cite{rudinPrinciplesMathematicalAnalysis1976} to $\mathcal{F}$ at interval $[\epsilon, l_1]$, there exists $l^\star \in [\epsilon, l_1]$ such that $\mathcal{L}_\epsilon(\hat{p}_\mathbf{x}, p_\mathbf{x}) = \mathcal{L}_{\Sigma^{l^\star}}(\hat{p}_\mathbf{x}, p_\mathbf{x}).$ Letting $\Sigma^\star = \Sigma^{l^\star}$, the proof is completed following from the evaluation in Proposition \ref{thm:evl_finite}.}

\section{Proof of Lemma \ref{lem:approx_onestepLog}}\label{app:lem:approx_onestepLog}
When $\epsilon = 0$, the result trivially follows from the definition of the expected logarithm score. \textcolor{blue}{When $\epsilon > 0$, define an error functional $\delta_\epsilon^N: \mathcal{P}_c \to \R$ by
    \[\delta_\epsilon^N(h) := \max\limits_{\|x_1-x_2\|\leq\epsilon, \max\{\|x_1\|, \|x_2\|\}  \leq N} \left|\log\frac{h(x_1)}{h(x_2)}\right|,\]
    where $N\in\R_+$, $x_1,x_2 \in \R^{d_x}$.  $\delta_\epsilon^N(h)$ is monotonically increasing regarding $N$ or $\epsilon$ because the feasible region for $(x_1, x_2)$ is enlarged when $N$ or $\epsilon$ increases.}

\textcolor{blue}{According to Theorem \ref{thm:existence}, $\exists \Sigma^\star$ s.t. $\mathcal{L}_\epsilon(\hat{p}_\mathbf{x}, p_\mathbf{x}) = \mathcal{L}_{\Sigma^\star}(\hat{p}_\mathbf{x}, p_\mathbf{x})$. Suppose $\Sigma^\star = \Sigma^{\kappa\epsilon} = \left\{A_{v}\right\}_{v\in \mathbb{Z}^{d_x}}$.
Since $p_{\mathbf{x}}(\cdot)$ is continuous and any cube $A_v$ is bounded, according to the intermediate value theorem, $\forall v\in\mathbb{Z}^{d_x}, \exists a_v \in A_v$ such that $p_{\mathbf{x}}(a_v)|A_v| = p_{\mathbf{x}}^{\Sigma^\star}(v)$, where $|A_v|$ denotes the Lebesgue volume of $A_v$. Then $\mathcal{L}_\epsilon(\hat{p}_{\mathbf{x}}, p_{\mathbf{x}})$ equals
$$
    \begin{aligned}
          & \sum_{v\in\mathbb{Z}^{d_x}} p_{\mathbf{x}}^{\Sigma^\star}(v)\log \left[\hat{p}_{\mathbf{x}}(a_v)\cdot|A_v|\right]                                                                   \\
        = & \sum_{v\in\mathbb{Z}^{d_x}} p_{\mathbf{x}}^{\Sigma^\star}(v)\log\left[|A_v|\right] + \sum_{v\in\mathbb{Z}^{d_x}} p_{\mathbf{x}}(a_v)|A_v|\log\left[\hat{p}_{\mathbf{x}}(a_v)\right] \\
        = & d_x\log(\kappa\epsilon) + \sum_{v\in\mathbb{Z}^{d_x}} p_{\mathbf{x}}(a_v)|A_v|\log\left[\hat{p}_{\mathbf{x}}(a_v)\right].
    \end{aligned}
$$
Notice that $\forall \tau \in (0, \epsilon/2)$, $\exists N_1 > 0$ such that
\begin{equation*}
    \left|\sum_{\|v\|> N_1/(\kappa\epsilon)} p_{\mathbf{x}}(a_v)|A_v|\log\left[\hat{p}_{\mathbf{x}}(a_v)\right]\right| < \tau.
\end{equation*}
Next, notice that $\mathrm{H}(p_{\mathbf{x}}||\hat{p}_{\mathbf{x}})$ is defined by an integration over $\R^{d_x}$, $\forall \tau \in (0, \epsilon/2)$, $\exists N_2 > 0$ such that
\begin{equation*}
    \left|\sum_{\|v\|> N_2/(\kappa\epsilon)}\int_{A_v}p_{\mathbf{x}}(x)\log \hat{p}_{\mathbf{x}}(x) \dd x\right|< \tau.
\end{equation*}
Let $N_3 = \max\{N_1, N_2\}$, there is
\begin{equation*}\label{eq:app5_2}
    \begin{aligned}
             & \left|\sum_{v\in\mathbb{Z}^{d_x}} p_{\mathbf{x}}(a_v)|A_v|\log\left[\hat{p}_{\mathbf{x}}(a_v)\right]+\mathrm{H}(p_{\mathbf{x}}||\hat{p}_{\mathbf{x}})\right|                                     \\
        =    & \left|\sum_{v\in\mathbb{Z}^{d_x}}\!\! p_{\mathbf{x}}(a_v)|A_v|\log\left[\hat{p}_{\mathbf{x}}(a_v)\right]\!-\!\!\int_{A_v}\!p_{\mathbf{x}}(x)\log\left[\hat{p}_{\mathbf{x}}(x)\right]\dd x\right| \\
        \leq & \left|\sum_{\|v\|> N_3/(\kappa\epsilon)} p_{\mathbf{x}}(a_v)|A_v|\log\left[\hat{p}_{\mathbf{x}}(a_v)\right] \right|+                                                                             \\
             & \left| \sum_{\|v\|\leq N_3/(\kappa\epsilon)} \int_{A_v}p_{\mathbf{x}}(x)\log\left(\frac{\hat{p}_{\mathbf{x}}(x)}{\hat{p}_{\mathbf{x}}(a_v)}\right)\dd x \right|+                                 \\
             & \left|\sum_{\|v\|> N_3/(\kappa\epsilon)}\int_{A_v}p_{\mathbf{x}}(x)\log \hat{p}_{\mathbf{x}}(x) \dd x\right|                                                                                     \\
        <    & \tau + \sum_{\|v\|\leq N_3/(\kappa\epsilon)}\left|\int_{A_v}p_{\mathbf{x}}(x)\dd x\right|\delta^{N_3}_{\kappa\epsilon}(\hat{p}_{\mathbf{x}})+ \tau                                               \\
        \leq & \delta^{N_3}_{\kappa\epsilon}(\hat{p}_{\mathbf{x}}) + 2\tau
    \end{aligned}
\end{equation*}
It follows that
\begin{equation}\label{eq:app5_1}
    |\mathcal{L}_\epsilon(\hat{p}_{\mathbf{x}}, p_{\mathbf{x}}) + \mathrm{H}(p_{\mathbf{x}}||\hat{p}_{\mathbf{x}})-d_x\log(\kappa\epsilon)|
    <\delta^{N_3}_{\kappa\epsilon}(\hat{p}_{\mathbf{x}})+2\tau.
\end{equation}
\eqref{eq:app5_1} is quite close to our objective except for the $d_x\log(\kappa\epsilon)$ term. Immediately,
\begin{equation}\label{eq:app5_2}
    \begin{aligned}
             & |\mathcal{L}_\epsilon(\hat{p}_{\mathbf{x}}, p_{\mathbf{x}}) + \mathrm{H}(p_{\mathbf{x}}||\hat{p}_{\mathbf{x}})-d_x\log(2\epsilon)|                                                       \\
        \leq & \left|\mathcal{L}_\epsilon(\hat{p}_{\mathbf{x}}, p_{\mathbf{x}}) +\mathrm{H}(p_{\mathbf{x}}||\hat{p}_{\mathbf{x}})-d_x\log(\kappa\epsilon)\right|+\left|d_x\log(\frac{2}{\kappa})\right| \\
        <    & \delta^{N_3}_{\kappa\epsilon}(\hat{p}_{\mathbf{x}})+2\tau + \left|d_x\log(\frac{2}{\kappa})\right|
    \end{aligned}
\end{equation}
The next goal is to find an upper bound for $\left|d_x\log(\frac{2}{\kappa})\right|$.
Guaranteed by the law of large numbers, given a sequence of samples $\{x^r\}_{r=1}^\infty$ with $x^r \overset{i.i.d}{\sim} p_{\mathbf{x}}$, there is
\begin{equation}\label{eq:upp_eps}
    \begin{aligned}
        \mathcal{L}_\epsilon(\hat{p}_{\mathbf{x}}, p_{\mathbf{x}})
        = & \lim_{R\to\infty} \frac{1}{R} \sum_{r=1}^{R} \log\left(\int_{\|x^r -s\|\leq \epsilon} \hat{p}_{\mathbf{x}}(s)\dd s \right) \\
        = & d_x \log(2\epsilon) + \lim_{R\to\infty} \frac{1}{R} \sum_{r=1}^{R} \log \hat{p}_{\mathbf{x}}(x^r_1),
    \end{aligned}
\end{equation}
where $\|x^r_1 - x^r\|\leq\epsilon$ s.t. $(2\epsilon)^{d_x}\hat{p}_{\mathbf{x}}(x^r_1) = \int_{\|x^r - s\|} \hat{p}_{\mathbf{x}}(s) \dd s$. Similarly, there is
\begin{equation}\label{eq:upp_sig}
    \mathcal{L}_{\Sigma^\star}(\hat{p}_{\mathbf{x}}, p_{\mathbf{x}}) = d_x\log(\kappa\epsilon) + \lim_{R\to\infty} \frac{1}{R} \sum_{r=1}^{R} \log \hat{p}_{\mathbf{x}}(x^r_2),
\end{equation}
where $x^r_2 \in A_{\Theta_{\Sigma^\star}(x^r)}$ s.t. $(\kappa\epsilon)^{d_x} \hat{p}_{\mathbf{x}}(x^r_2) = \hat{p}_{\mathbf{x}}^{\Sigma^\star}(\Theta_{\Sigma^\star}(x^r))$. Since \eqref{eq:upp_eps} and \eqref{eq:upp_sig} are equal, one has
\begin{equation*}
    d_x\log(\frac{2}{\kappa}) \!=\! \lim_{R\to\infty} \frac{1}{R} \!\sum_{r=1}^{R} \log \hat{p}_{\mathbf{x}}(x^r_2) \!-\! \lim_{R\to\infty} \frac{1}{R}\! \sum_{r=1}^{R} \log \hat{p}_{\mathbf{x}}(x^r_1).
\end{equation*}
Notice that $\forall \tau \in (0, \epsilon/2)$, $\exists N_4 > 0$ s.t.
\begin{equation*}
    \begin{aligned}
        \tau > & \left|\int_{\|x\|> N_4} p_{\mathbf{x}}(x) \log\left(\frac{\int_{\|x -s\|\leq \epsilon} \hat{p}_{\mathbf{x}}(s)\dd s}{(2\epsilon)^{d_x}} \right) \dd x \right|                   \\
        =      & \left| \lim_{R\to\infty} \frac{1}{R}\! \sum_{r=1}^{R} \mathbb{I}_{\|x^r\|>N_4} \cdot \log  \hat{p}_{\mathbf{x}}(x^r_1) \right|,                                                 \\
        \tau > & \left|\sum_{\|v\|> N_4/(\kappa\epsilon)} p_{\mathbf{x}}^{\Sigma^\star}(v) \log\left(\frac{\hat{p}_{\mathbf{x}}^{\Sigma^\star}(v)}{(\kappa\epsilon)^{d_x}} \right) \dd x \right| \\
        =      & \left| \lim_{R\to\infty} \frac{1}{R}\! \sum_{r=1}^{R} \mathbb{I}_{\|x^r\|>N_4} \cdot \log  \hat{p}_{\mathbf{x}}(x^r_2) \right|.
    \end{aligned}
\end{equation*}
Therefore, $\left|d_x\log(\frac{2}{\kappa})\right|$ equals
\begin{equation}\label{eq:kappa}
    \begin{aligned}
             & \left|\lim_{R\to\infty} \frac{1}{R} \!\sum_{r=1}^{R} \log \hat{p}_{\mathbf{x}}(x^r_2) \!-\! \lim_{R\to\infty} \frac{1}{R}\! \sum_{r=1}^{R} \log \hat{p}_{\mathbf{x}}(x^r_1) \right| \\
        \leq & \left| \lim_{R\to\infty} \frac{1}{R}\! \sum_{r=1}^{R} \mathbb{I}_{\|x^r\|>N_4} \cdot \log  \hat{p}_{\mathbf{x}}(x^r_1) \right| +                                                    \\
             & \left| \lim_{R\to\infty} \frac{1}{R}\! \sum_{r=1}^{R} \mathbb{I}_{\|x^r\|>N_4} \cdot \log  \hat{p}_{\mathbf{x}}(x^r_2) \right|  +                                                   \\
             & \lim_{R\to\infty}\frac{1}{R}\sum\limits_{r=1}^R\mathbb{I}_{\|x^r\|\leq N_4}\cdot \left|\log \hat{p}_{\mathbf{x}}(x^r_1)\!-\!\log \hat{p}_{\mathbf{x}}(x^r_2)\right|                 \\
        <    & \delta_{(\kappa+1)\epsilon}^{N_4}(\hat{p}_{\mathbf{x}})+\epsilon.
    \end{aligned}
\end{equation}
Let $\bar{N} = \max\{N_3, N_4\}$, we have
\begin{equation*}
    |\mathcal{L}_\epsilon(\hat{p}_{\mathbf{x}}, p_{\mathbf{x}}) + \mathrm{H}(p_{\mathbf{x}} || \hat{p}_{\mathbf{x}})-d_x\log(2\epsilon)| < 2\left(\delta^{\bar{N}}_{(\kappa+1)\epsilon}(\hat{p}_{\mathbf{x}})+\epsilon\right).
\end{equation*}
Consider another functional operator $\rho: \mathcal{P}_c \to \R$ as the maximum value of the solution set of an inequality, i.e.,
\[\rho(h):=\max_{z\in\R_{+}}\left\{z:\left|d_x\log(\frac{2}{z})\right|\leq \delta^{\bar{N}}_{(z+1)\epsilon}(h) + \epsilon \right\}.\]
\eqref{eq:kappa} ensures that $\kappa$ belongs to the solution set. Notice that $\delta^{\bar{N}}_{z\epsilon}(\hat{p}_{\mathbf{x}})$ is upper bounded regarding $z$ while $\left|d_x\log(\frac{2}{z})\right|$ is not, one has the solution set is upper bounded and $\rho(\hat{p}_{\mathbf{x}})$ is well-defined.  It follows that
\begin{equation*}
    \left|d_x\log(\frac{2}{\kappa})\right| \leq \delta^{\bar{N}}_{\kappa\epsilon}(\hat{p}_{\mathbf{x}})+\epsilon \leq \delta^{\bar{N}}_{\rho(\hat{p}_{\mathbf{x}})\epsilon}(\hat{p}_{\mathbf{x}}) + \epsilon.
\end{equation*}
Since $\log \hat{p}_{\mathbf{x}}(\cdot)$ is uniformly continuous over the bounded set $\{x \mid \|x\|\leq \bar{N}\}$, we have $\delta^{\bar{N}}_{\rho(\hat{p}_{\mathbf{x}})\epsilon}(\hat{p}_{\mathbf{x}}) = \mathcal{O}(\epsilon)$, and the proof is completed.}

\section{Proof of Theorem \ref{thm:approx_elog_traj}}\label{app:thm:approx_elog_traj}
When $\epsilon = 0$, the result trivially follows from the definition of the expected logarithm score. When $\epsilon > 0$, let $q_{x_{1:k-1}}(\cdot)= p_{\mathbf{x}_k \mid \mathbf{x}_{1:k-1}}(\cdot \mid x_{1:k-1})$ and $\hat{q}_{x_{1:k-1}}(\cdot)= \hat{p}_{\mathbf{x}_k \mid \mathbf{x}_{1:k-1}}(\cdot \mid x_{1:k-1})$, it follows follows from the \textcolor{blue}{the chain rules for joint entropy and relative entropy \cite[pp. 22-24]{coverElementsInformationTheory2006} that}
\begin{equation*}
    \begin{aligned}
          & \left|\bar{\mathcal{L}}_\epsilon(\hat{p}_{\mathbf{x}_{1:T}}, p_{\mathbf{x}_{1:T}}) \!-\! \left\{d_x\log(2\epsilon)-\frac{1}{T}\mathrm{H}(p_{\mathbf{x}_{1:T}} || \hat{p}_{\mathbf{x}_{1:T}})\right\}\right| \\
        = & \bigg| \frac{1}{T}\!\sum_{k=1}^{T} \E_{x_{1:k-1}}\mathcal{L}_\epsilon\left(\hat{q}_{x_{1:k-1}}, q_{x_{1:k-1}}\right)                                                                                        \\
          & - \left\{d_x\log(2\epsilon)-\frac{1}{T}\sum_{k=1}^{T}\mathrm{H}(p_{\mathbf{x}_k \mid \mathbf{x}_{1:k-1}}|| \hat{p}_{\mathbf{x}_k \mid \mathbf{x}_{1:k-1}})\right\} \bigg|,
    \end{aligned}
\end{equation*}
which can be upper bounded by
\begin{equation*}
    \begin{aligned}
         & \frac{1}{T}\!\sum_{k=1}^{T} \E_{x_{1:k-1}} \bigg|\mathcal{L}_\epsilon\left(q_{x_{1:k-1}}, q_{x_{1:k-1}}\right) - d_x\log(2\epsilon) +                        \\
         & \mathrm{H}\left(p_{\mathbf{x}_k \mid \mathbf{x}_{1:k-1}}(\cdot | x_{1:k-1})|| \hat{p}_{\mathbf{x}_k \mid \mathbf{x}_{1:k-1}}(\cdot | x_{1:k-1})\right)\bigg|
        \overset{(i)}{=}   \mathcal{O}(\epsilon),
    \end{aligned}
\end{equation*}
where $(i)$ holds because Lemma \ref{lem:approx_onestepLog} ensures each term is $\mathcal{O}(\epsilon)$, the average of finite sum is also of the scale $\mathcal{O}(\epsilon)$.

\section{Proof of Theorem \ref{thm:converge}} \label{app:thm:converge}
According to the assumption that the process noises are i.i.d, $\bar{\mathcal{L}}_\epsilon(\hat{p}_{\mathbf{x}_{1:T}}, x_{1:T})$ equals
\begin{equation*}
    \begin{aligned}
                          & \frac{1}{T}\!\sum_{k=1}^{T} \mathcal{L}_\epsilon\left(\hat{p}_{\mathbf{x}_k \mid \mathbf{x}_{1:k-1}}(\cdot \mid x_{1:k-1}), x_k\right)                                                                                                           \\
        \overset{(i)}{=}  & \frac{1}{T}\!\sum_{k=1}^{T} \mathcal{L}_\epsilon\left(\hat{p}_{\mathbf{x}_k \mid \mathbf{x}_{k-1}}(\cdot \mid x_{k-1}), x_k\right)                                                                                                               \\
        \overset{(ii)}{=} & \frac{1}{T}\!\sum_{k=1}^{T} \mathcal{L}_\epsilon\left(\hat{p}_{\mathbf{w}}(\cdot), x_k - f(x_{k-1}) \right)  \overset{(iii)}{=}                          \frac{1}{T}\!\sum_{k=1}^{T} \mathcal{L}_\epsilon\left(\hat{p}_{\mathbf{w}}, w_k\right), \\
    \end{aligned}
\end{equation*}
where $(i)$ holds because when the process noises $\{\mathbf{w}_k\}_{k=1}^T$ are independent, $\{\mathbf{x}_k\}_{k=1}^T$ is a Markov process, thus $p_{\mathbf{x}_k \mid \mathbf{x}_{1:k-1}} = p_{\mathbf{x}_k \mid \mathbf{x}_{k-1}}$; \textcolor{blue}{$(ii)$ and $(iii)$ follows from the assumption on the predictors where the prediction is reduced to predicting the noises $\mathbf{w}_k\overset{i.i.d}{\sim}p_{\mathbf{w}}$ by $\hat{p}_{\mathbf{w}}$. Then we can derive $\bar{\mathcal{L}}_{\epsilon}(\hat{p}_{\mathbf{x}_{1:T}}, p_{\mathbf{x}_{1:T}})$ as
\begin{equation*}
    \frac{1}{T}\sum_{k=1}^{T} \mathbb{E}\left\{ \mathcal{L}_\epsilon\left(\hat{p}_{\mathbf{w}}, w_k\right)\right\}  = \mathcal{L}_\epsilon\left(\hat{p}_{\mathbf{w}}, p_{\mathbf{w}}\right).
\end{equation*}
Ensured by the strong law of large numbers, one has
\begin{equation*}
    \begin{aligned}
        \lim\limits_{T\to\infty}\bar{\mathcal{L}}(\hat{p}_{\mathbf{x}_{1:T}}, x_{1:T}) = & \lim\limits_{T\to\infty}\frac{1}{T}\sum_{k=1}^{T} \mathcal{L}_\epsilon\left(\hat{p}_{\mathbf{w}}, w_k\right) \\
                                                                                         & \xrightarrow{a.s.}   \mathcal{L}_\epsilon\left(\hat{p}_{\mathbf{w}}, p_{\mathbf{w}}\right).
    \end{aligned}
\end{equation*}
Finally,} if $\E_{w\sim p_{\mathbf{w}}} \mathcal{L}_\epsilon(\hat{p}_\mathbf{w}, w)^2 < \infty$, it follows that
\begin{equation*}
    \begin{aligned}
          & \var\{\bar{\mathcal{L}}_\epsilon(\hat{p}_{\mathbf{x}_{1:T}}, x_{1:T})\}= \var\left\{\frac{1}{T}\!\sum_{k=1}^{T} \mathcal{L}_\epsilon\left(\hat{p}_{\mathbf{w}}, w_k\right)\right\} \\
        = & \frac{1}{T} \left\{\E_{w\sim p_{\mathbf{w}}} \mathcal{L}_\epsilon(\hat{p}_\mathbf{w}, w)^2 - \mathcal{L}_\epsilon(\hat{p}_\mathbf{w}, p_\mathbf{w})^2\right\}.
    \end{aligned}
\end{equation*}
Ensured by the Chebyshev inequality, the converging speed is $\mathcal{O}_p(\frac{1}{\sqrt{T}})$, i.e., $\forall \delta > 0$, there is
\begin{equation*}
    \pr\left\{|\bar{\mathcal{L}}_\epsilon(\hat{p}_{\mathbf{x}_{1:T}}, x_{1:T}) - \bar{\mathcal{L}}_\epsilon(\hat{p}_{\mathbf{x}_{1:T}}, p_{\mathbf{x}_{1:T}})| \geq \delta \right\} = \mathcal{O}(\frac{1}{\sqrt{T}}).
\end{equation*}
\end{document}